\documentclass[10pt, twoside]{article}

\usepackage[tbtags]{amsmath}

\usepackage{amsfonts}
\usepackage{amssymb}
\usepackage{latexsym}

\DeclareMathSymbol{\vecarrow}{\mathord}{letters}{"7E}

\newcommand{\starMP}{*_{\scriptscriptstyle MC}}
\newcommand{\starM}{*_{\scriptscriptstyle M}}

\newcommand{\starP}{*_{\scriptscriptstyle C}}

\newcommand{\cdotSy}{\cdot_{\scriptscriptstyle Sy}}

\newcommand{\fett}[1]{\mbox{\boldmath$#1$}} 

\newcommand{\iu}{{\mathrm{i}}}

\newcommand{\beq}{\begin{equation}}
\newcommand{\eeq}{\end{equation}}

\begin{document}
\makeatletter
\title{Deformed Geometric Algebra and Supersymmetric Quantum Mechanics}
\author{Peter Henselder\footnote{peter.henselder@uni-dortmund.de}\\
Fachbereich Physik, Universit\"at Dortmund\\
44221 Dortmund}

\maketitle

\begin{abstract}
Deforming the algebraic structure of geometric algebra on the
phase space with a Moyal product leads naturally to supersymmetric
quantum mechanics in the star product formalism.
\end{abstract}

\section{Introduction}
\qquad Quantum mechanics has a natural description on the phase
space in the star product formalism \cite{Bayen1,Zachos3}.
Moreover spin and relativistic quantum mechanics can also be
described in the star product formalism if one deforms
pseudoclassical mechanics \cite{Berezin1} with a fermionic star
product. This was done in \cite{Deform3} and it was also shown
that such a fermionic sector relates in supersymmetric quantum
mechanics the supersymmetric partner systems.

One might then wonder which physical status pseudoclassical
mechanics and its deformed version actually have. This question
was solved in \cite{Doran2} where the relation of pseudoclassical
mechanics and geometric algebra was established (for a
comprehensive discussion of geometric algebra see for example
\cite{Doran1}). It becomes then clear that the fermionic sector
describes the geometric structure of the phase space in a
multivector formalism. Furthermore the fermionic star product
corresponds to the geometric product of geometric algebra that
deforms Grassmann calculus into Clifford calculus \cite{Deform6}.

Geometric algebra in its superanalytic formulation with the
Clifford star product as the geometric product can then be used to
describe the Hamilton formalism on the phase space. Moreover it
appears natural to combine the fermionic star product that
describes the geometric structure with the bosonic Moyal star
product that makes this structure noncommutative. The result is a
deformed, noncommutative version of geometric algebra. For
geometric algebra on three-space the transition to
noncommutativity induces an extra term that splits the system in a
version with spin up and one with spin down, i.e.\
noncommutativity in geometric algebra on three-space transforms
the Schr\"odinger Hamiltonian into the Pauli Hamiltonian \cite{Deform6}.
As will be shown below the transition to noncommutativity for
geometric algebra on the phase space leads similarly to a split
into two supersymmetric partner systems.

One can see the appearance of geometric algebra structures on the
phase space also in a different way: Just as the factorization of
the Klein-Gordon equation exhibits in Dirac theory the Clifford
structure of space-time \cite{Hestenes1}, the factorization of a
Hamilton function into supercharges exhibits the Clifford
structure of the phase space.

\section{Geometric Algebra on the phase space}
\setcounter{equation}{0}\label{BFosc}
Geometric algebra was first used on the phase space to describe
the Hamilton formalism in \cite{Hestenes9}. We will here restrict
to the simplest case of a flat two-dimensional phase space and use
the superanalytic formulation of geometric algebra. A point in the
phase space is a vector or supernumber of Grassmann grade one:
\begin{equation}
\fett{z}=z^i\fett{\zeta}_i=q\fett{\eta}+p\fett{\rho},
\label{xphasenraum}
\end{equation}
where the Grassmann variables $\fett{\zeta}_1=\fett{\eta}$ and
$\fett{\zeta}_2=\fett{\rho}$ are the basis vectors of the two
dimensional vector space. On this vector space the Clifford star
product of two multivectors $A$ and $B$ is given by
\begin{equation}
A\starP B=A\,\exp\left[\eta_{ij}
\frac{\overleftarrow{\partial}}{\partial\fett{\zeta}_i}
\frac{\overrightarrow{\partial}}{\partial\fett{\zeta}_j}\right]\,B,
\label{internstarP}
\end{equation}
where $\eta_{ij}=\mathrm{diag}(1,1)$ is the euclidian metric on
the vector space. Furthermore one has a closed two-form
\begin{equation}
\Omega=\frac{1}{2}\Omega_{ij}\fett{\zeta}^i\fett{\zeta}^j
=\fett{\eta}\fett{\rho}=\fett{d}q\fett{d}p,\label{OmegaDef}
\end{equation}
where $\Omega_{ij}$ is a non-degenerate, antisymmetric matrix
\cite{McDuff} and $\fett{d}=\fett{\zeta}^i\frac{\partial}{\partial
z_i}=\fett{\eta}\frac{\partial}{\partial
q}+\fett{\rho}\frac{\partial}{\partial p}$ is the nabla operator.

The euclidian scalar product of two vectors
$\fett{a}=a^i\fett{\zeta}_i$ and $\fett{b}=b^i\fett{\zeta}_i$ is
given by the scalar part of their star product (\ref{internstarP}),
i.e.\ $\fett{a}\cdot\fett{b}=\langle\fett{a}\starP
\fett{b}\rangle_0=a^ib_i\eta_{ij}$ and with the two-form $\Omega$ 
the symplectic scalar product is given by
\begin{equation}
\fett{a}\cdotSy\fett{b}=(\fett{ba})\cdot\Omega
=\fett{a}\cdot(\Omega\cdot\fett{b})=a^i\Omega_{ij}b^j.\label{cdotSy}
\end{equation}
Furthermore one can map with $\Omega$ a vector in a one-form
according to $\fett{a}^{\flat}=\fett{a}\cdot\Omega$. The inverse
map of a one-form into a vector can be described with the bivector
\begin{equation}
\mathtt{J}=\frac{1}{2}J^{ij}\fett{\zeta}_i\fett{\zeta}_j
\end{equation}
so that the vector corresponding to a one-form
$\fett{\omega}=\omega_i\fett{\zeta}^i$ is given by
$\fett{\omega}^{\natural}=\mathtt{J}\cdot\fett{\omega}$. The map
$\natural$ should be inverse to $\flat$, from which
$J^{ij}=(\Omega_{ij}^{-1})^T=\Omega^{ji}$ follows. The Hamilton
equations can then be written as:
\begin{equation}
\dot{\fett{z}}=\fett{d}^{\natural}H
\end{equation}
and for the Poisson bracket one has
\begin{equation}
\{F,G\}_{PB}=F\,\overleftarrow{\fett{d}}\cdotSy\overrightarrow{\fett{d}}\, G
=J^{ab}\frac{\partial F}{\partial x^a}\frac{\partial G}{\partial
x^b}.
\end{equation}

\section{Star-Factorization of the Hamilton Function}
\setcounter{equation}{0}
A Hamilton function can be written as the square of the vector
\begin{equation}
\fett{w}=W(q)\fett{\eta}+p\fett{\rho}, \label{wdef}
\end{equation}
where $W(q)$ is the Superpotential, one has then 
\begin{equation}
H=\frac{1}{2}\fett{w}\starP\fett{w}
=\frac{1}{2}\fett{w}\cdot\fett{w}
=\frac{1}{2}\left[p^2+W^2(q)\right]
\end{equation}
and in holomorphic coordinates $B=\frac{1}{\sqrt{2}}(W(q)+\iu p)$,
$\bar{B}=\frac{1}{\sqrt{2}}(W(q)-\iu p)$ and
$\fett{f}=\frac{1}{\sqrt{2}}(\fett{\eta}+\iu\fett{\rho})$,
$\bar{\fett{f}}=\frac{1}{\sqrt{2}}(\fett{\eta}-\iu\fett{\rho})$
one obtains
\begin{equation}
\fett{w}=B\bar{\fett{f}}+\bar{B}\fett{f}=\fett{Q}_++\fett{Q}_-
\label{wholodef}
\end{equation}
and $H=B\bar{B}$.

Up to now the coefficients were commuting quantities, but one can
go over to the noncommutative or quantum case by demanding that the
coefficients have to be multiplied by the Moyal product
\begin{equation}
f\starM g
=f\,\exp\left[\frac{\iu\hbar}{2}\left(\frac{\overleftarrow{\partial}}
{\partial q}\frac{\overrightarrow{\partial}}{\partial p}
-\frac{\overleftarrow{\partial}}{\partial p}
\frac{\overrightarrow{\partial}}{\partial q} \right)\right]\,g.
\end{equation}
In this case the square of $\fett{w}$ is no longer a scalar, but
one has an bivector valued extra term
\begin{eqnarray}
H_S=\frac{1}{2}\fett{w}\starMP\fett{w} &=&\frac{1}{2}
\big[(W(q)\starM W(q))(\fett{\eta}\starP\fett{\eta}) +(W(q)\starM
p)(\fett{\eta}\starP\fett{\rho}) \nonumber\\
&&\quad +(p\starM W(q))(\fett{\rho}\starP
\fett{\eta})+(p\starM p)(\fett{\rho}\starP\fett{\rho})\big]\\
&=&\frac{1}{2}\left[p^2+W^2(q)\right]+\frac{\hbar}{2}\frac{\partial
W(q)}{\partial q}\frac{1}{\iu}\fett{\eta\rho}. \label{HSdef}
\end{eqnarray}
The next thing one has to notice is that $\fett{\eta}$,
$\fett{\rho}$ and $-\iu\fett{\eta\rho}$ fulfill under the Clifford
star product the Pauli algebra, i.e.\ one has for the star
commutators $\left[A,B\right]_{\starP} =A\starP B-B\starP A$ and
anticommutators $\{A,B\}_{\starP} =A\starP B+B\starP A$ of these
real basis elements of the two dimensional Clifford algebra:
\begin{eqnarray}
&&\left[\fett{\eta},\fett{\rho}\right]_{\starP}=2\fett{\eta\rho},
\qquad
\left[\fett{\eta},-\iu\fett{\eta\rho}\right]_{\starP}=-2\iu\fett{\rho},
\qquad
\left[\fett{\rho},-\iu\fett{\eta\rho}\right]_{\starP}=2\iu\fett{\eta}\\
&\mathrm{and}& \{\fett{\eta},\fett{\eta}\}_{\starP}=
\{\fett{\rho},\fett{\rho}\}_{\starP}=
\{-\iu\fett{\eta\rho},-\iu\fett{\eta\rho}\}_{\starP}=2,
\end{eqnarray}
while the other star commutators and star anticommutators vanish.
This means that $\fett{\eta}$, $\fett{\rho}$ and
$-\iu\fett{\eta\rho}$ would be represented in a tuple
representation by the Pauli matrices, so that $H_S$ is the
supersymmetric Hamiltonian. Furthermore for the holomorphic basis
vectors $\fett{f}=\frac{1}{\sqrt{2}}(\fett{\eta}+\iu\fett{\rho})$
and
$\bar{\fett{f}}=\frac{1}{\sqrt{2}}(\fett{\eta}-\iu\fett{\rho})$
one has the tuple representation
\begin{equation}
\frac{1}{\sqrt{2}}\fett{f}\cong \left(
\begin{array}{cc}
  0 & 1 \\
  0 & 0 \\
\end{array}\right)\quad\mathrm{and}\quad
\frac{1}{\sqrt{2}}\bar{\fett{f}}\cong \left(
\begin{array}{cc}
  0 & 0 \\
  1 & 0 \\
\end{array}\right).
\end{equation}

The two eigen-multivectors of $-\iu\fett{\eta\rho}$ are
$\pi^{(C)}_{\pm}=\frac{1}{2}(1\mp\iu\fett{\eta\rho})$, i.e.\ for
these multivectors one has
\begin{equation}
-\iu\fett{\eta\rho}\starP\pi^{(C)}_{\pm}=\pm\pi^{(C)}_{\pm}.
\end{equation}
In the star product formalism these multivectors are fermionic
Wigner functions and as such they are projectors:
\begin{equation}
\pi^{(C)}_{\pm}\starP\pi^{(C)}_{\pm}=\pi^{(C)}_{\pm}\quad
\mathrm{and}\quad \pi^{(C)}_+\starP\pi^{(C)}_-=
\pi^{(C)}_-\starP\pi^{(C)}_+=0,\label{proje}
\end{equation}
while in geometric algebra these multivectors are related to
spinors \cite{Francis1}. The holomorphic basis vectors
$\frac{1}{\sqrt{2}}\fett{f}$ and
$\frac{1}{\sqrt{2}}\bar{\fett{f}}$ serve here as lowering and
raising operators, i.e.\ one has
\begin{equation}
\bar{\fett{f}}\starP\pi_{+}^{(C)}\starP\fett{f}=2\pi_{-}^{(C)}
\quad\mathrm{and}\quad \fett{f}\starP\pi_{-}^{(C)}\starP
\bar{\fett{f}}=2\pi_{+}^{(C)},
\end{equation}
while the other combinations give zero.

With the multivectors $\pi_{\pm}^{(C)}$ the supersymmetric
Hamilton function (\ref{HSdef}) can then be written as
\begin{eqnarray}
H_S&=&\frac{1}{2}\left[p^2+W^2(q)-\hbar\frac{\partial
W(q)}{\partial q}\right]
\left(\frac{1}{2}-\frac{\iu}{2}\fett{\eta\rho}\right)\nonumber\\
&&+\frac{1}{2}\left[p^2+W^2(q)+\hbar\frac{\partial W(q)}{\partial
q}\right]
\left(\frac{1}{2}+\frac{\iu}{2}\fett{\eta\rho}\right)\\
&=&H_1\pi_+^{(C)}+H_2\pi_-^{(C)}.
\end{eqnarray}
From (\ref{proje}) it is then clear that the Moyal-Clifford star
eigenfunctions of $H_S$ are a product of $\pi_{+}^{(C)}$ and Moyal
star eigenfunctions of $H_1$ or products of $\pi_{-}^{(C)}$ and
Moyal star eigenfunctions of $H_2$. The Moyal star eigenfunctions
for supersymmetric partner potentials were for examples discussed
in \cite{Curtright}.

The vectors $\fett{Q}_{\pm}$ defined in (\ref{wholodef}) fulfill
\begin{equation}
\fett{Q}_{\pm}\starMP\fett{Q}_{\pm}=0,\quad
\fett{Q}_{-}\starMP\fett{Q}_{+}=H_1\pi_+^{(C)},\quad
\fett{Q}_{+}\starMP\fett{Q}_{-}=H_2\pi_-^{(C)} \label{Qrelas}
\end{equation}
so that $H_S$ can be written as
\begin{equation}
H_S=\frac{1}{2}\{\fett{Q}_+,\fett{Q}_-\}_{\starMP},
\end{equation}
and with (\ref{Qrelas}) one has $\left[\fett{Q}_+,H_S\right]_{\starMP}
=\left[\fett{Q}_-,H_S\right]_{\starMP}=0$. Defining eventually
\begin{equation}
\fett{Q}_1=\fett{Q}_++\fett{Q}_-\quad\mathrm{and}\quad
\fett{Q}_2=-\iu(\fett{Q}_+-\fett{Q}_-)
\end{equation}
the supersymmetric Hamilton function factorizes as
\begin{equation}
H_S=\frac{1}{2}\fett{Q}_1\starMP\fett{Q}_1=\frac{1}{2}
\fett{Q}_2\starMP\fett{Q}_2.
\end{equation}


\begin{thebibliography}{9999}

\bibitem{Bayen1} F.\ Bayen, M.\ Flato, C.\ Fronsdal, A.\ Lichnerowicz and
D.\ Sternheimer, Ann. Phys. {\bf 111} (1978) 61-110, 111-151.

\bibitem{Zachos3} C.\ Zachos,
Int. J. Mod. Phys. {\bf A17} (2002) 297-316.

\bibitem{Berezin1} F.\,A.\ Berezin and M.\,S.\ Marinov,
Ann. Phys. {\bf 104} (1977) 336-362.

\bibitem{Deform3} A.\,C.\ Hirshfeld and P.\ Henselder,
Ann. Phys. {\bf 302} (2002) 59-77.

\bibitem{Doran2} A.\ Lasenby, C.\ Doran and S.\ Gull,
J. Math. Phys. {\bf 34} (8) (1993) 3683-3712.

\bibitem{Doran1} C.\ Doran and A.\ Lasenby,
Geometric Algebra for Phyicists, Cambridge University Press
(2003).

\bibitem{Deform6} P.\ Henselder, A.\,C.\ Hirshfeld and T.\ Spernat,
Ann. Phys. {\bf 317} (2005) 107-129.

\bibitem{Hestenes1} D.\ Hestenes,
Space-Time Algebra, Gordon and Breach (1966).

\bibitem{Hestenes9} D.\ Hestenes, {\it Hamiltonian Mechanics with
Geometric Calculus}, in: Z. Oziewicz et al. (eds.), {Spinors,
Twistors, Clifford Algebras and Quantum Deformation} Kluwer
Dordercht/Boston (1993) 203-214.

\bibitem{McDuff} D.\ McDuff and D.\ Salamon, Introduction to
Symplectic Topology, Clarendon Press (1995).

\bibitem{Francis1} M.\,R.\ Francis and A.\ Kosowsky,
Ann. Phys. {\bf 317} (2005) 383-409.

\bibitem{Curtright} T.\ Curtright, D.\ Fairlie and C.\ Zachos,
Phys. Rev. D{\bf 58}(1998)025002.

























\end{thebibliography}
\end{document}